\documentclass[amsmath,amssymb,12pt]{revtex4-1}
\usepackage[utf8]{inputenc} 
\usepackage{graphicx}
\usepackage{float}
\usepackage{natbib}
\usepackage{wrapfig}
\usepackage[english]{babel}
\newcommand{\del}{\partial}
\begin{document}

\title{Complexity reduction in the 3D Kuramoto model}

\author{Ana Elisa D.  Barioni}

\author{Marcus A. M. de Aguiar }

\affiliation{Instituto de F\'isica `Gleb Wataghin', Universidade Estadual de Campinas, Unicamp 13083-970, Campinas, SP, Brazil}

\begin{abstract}
	
The dynamics of large systems of coupled oscillators is a subject of increasing importance with prominent applications
in several areas such as physics and biology. The Kuramoto model, where a set of oscillators move around a circle
representing their phases, is a paradigm in this field, exhibiting a continuous transition between disordered and
synchronous motion. Reinterpreting the oscillators as rotating unit vectors, the model was extended to allow vectors to
move on the surface of D-dimensional spheres, with $D=2$ corresponding to the original model. It was shown that the
transition to synchronous dynamics was discontinuous for odd D, raising a lot of interest. Inspired by results in 2D,
Ott et al \cite{chandra2019complexity} proposed an ansatz for density function describing the oscillators and derived
equations for the ansatz parameters, effectively reducing the dimensionality of the system. Here we take a different
approach for the 3D system and construct an ansatz based on spherical harmonics decomposition of the distribution
function. Our result differs significantly from that proposed in \cite{chandra2019complexity} and leads to similar but
simpler equations determining the dynamics of the order parameter. We derive the phase diagram of equilibrium solutions
for several distributions of natural frequencies and find excellent agreement with simulations. We also compare the
dynamics of the order parameter with numerical simulations and with the previously derived equations, finding good
agreement in all cases. We believe our approach can be generalized to higher dimensions and help to achieve  complexity
reduction in other systems of equations.

\end{abstract}

\maketitle

% %%%%%%%%%%%%%%%%%%%%%%%%%%%%%%%%%%%%%%%%%%%%%%%%%%%%
% %%%%%%%%%%%%%%%%%%%%%%%%%%%%%%%%%%%%%%%%%%%%%%%%%%%%
\section{Introduction}

The study of synchronization between elements of a system has a long history that dates back to Huygens
\cite{oliveira2014huygens}. Over the past years several examples of synchronization have been found in different
physical systems, from coupled metronomes \cite{Pantaleone2002} and electochemical oscillators \cite{Kiss2008} to neural
networks \cite{Novikov2014}, motivating the development of new mathematical methods to understand their global behavior.
One aspect of great importance is the transition from disordered to synchronized motion as the coupling intensity between the
elements increases.

In 1975 Yoshiki Kuramoto  proposed a simple model of $N$ coupled oscillators that could be solved analytically in the limit
where $N$ goes to infinity \cite{Kuramoto1975,Kuramoto1984}. The oscillators (or particles) were 
described by phases $\theta_i$ and natural frequencies $\omega_i$, selected from a symmetric distribution $g(\omega)$.
The oscillators interact according to the equations 
\begin{equation}
	\dot{\phi}_i = \omega_i + \frac{K}{N} \sum_{j=1}^N \sin{(\phi_j-\phi_i)}.
	\label{kuramoto}
\end{equation}
where $K$ is the coupling strength and $i = 1, ..., N$.  The complex order parameter
\begin{equation}
	z = p e^{i \psi} \equiv \frac{1}{N} \sum_{i=1}^N e^{i\theta_i}
	\label{paraord}
\end{equation}
measures the degree of phase synchronization of the particles: disordered motion implies $p \approx 0$ and coherent
motion $p \approx 1$. Kuramoto showed that the onset of sychronization could be described, in equilibrium, as a second
order phase transition, where $p$ remains very small for $0 < K < K_c$ and increases as $p = c\sqrt{K-K_c}$ for $
K>K_c$. These results  opened the way to large number of modifications and generalizations of his model, including
different types of coupling functions \cite{hong2011kuramoto,yeung1999time,breakspear2010generative}, introduction of
networks of connections (so that not all oscillators are connected to each other) \cite{Rodrigues2016,Joyce2019},
different distributions of the oscillator's natural frequencies (including frequencies proportional to the number of
connections, leading to explosive synchronization) \cite{Gomez-Gardenes2011,Ji2013}, inertial terms
\cite{Acebron2005,dorfler2011critical,olmi2014hysteretic} and external periodic driving forces
\cite{Childs2008,moreira2019global,moreira2019modular}. More recently, interest has been shifted to understand
oscillations in larger dimensions. It has been shown, in particular, that the natural extension of the Kuramoto model to
more dimensions exhibit first order phase transitions in odd dimensions and second order transitions in even dimensions
\cite{chandra2019continuous}.

In terms of mathematical methods, a breakthrough insight was given by Ott and Antonsen \cite{Ott2008}, who considered
the continuity equation satisfied by the oscillators and proposed an ansatz for their distribution on the unit circle
involving only two time dependent parameters. In the special case where the natural frequencies of the oscillators were
given by a Lorentzian distribution, the differential equations for the ansatz parameters could be translated into
equations for the order parameter $z$, allowing the system to be studied through two simple equations. This dimensional
reduction permitted, in particular, the construction of complete bifurcation diagrams when the oscillators are driven by
periodic forces \cite{Childs2008}. External driving forces acting only  on subsets of oscillators
\cite{moreira2019global} were also used to probe modular structures in neural networks \cite{moreira2019modular}.

The possibility of describing a large system of $N$ coupled particles by a small set of time-dependent parameters was
pointed out by Watanabe and Strogatz \cite{watanabe1994constants,Goebel1995} and further developed in \cite{Marvel2009}.
They showed that the trajectory of the i-th Kuramoto oscillator can be described by the Moebius transformation 
\begin{equation}
	z_i(t) = e^{ i \sigma(t)} \frac{a(t) - z_i(0)}{1-a^*(t)  z_i(0)}
\end{equation}
where $z_i=e^{i \phi_i}$, $\sigma$ is real and $a$ complex. The Moebius transformation was generalized by Tanaka
\cite{Tanaka2014} who extended it to multidimensional real of complex variables. For complex variables, attractive
coupling between particles (similar to the Kuramoto model) and multivariate Lorentzian distribution of natural
frequencies, equations of motion for the order parameter similar to the Stuart-Landau equation could be derived. For
real variables, corresponding, for instance, to the Kuramoto model in 3 dimensions, an equation for the order parameter
could only be obtained for the case of identical oscillators.

In this work we consider the 3-dimensional Kuramoto model proposed in \cite{chandra2019continuous}. In this model the
phases of the 2D oscillators are  re-interpreted  as unit vectors rotating on the surface of a circle and extended to
unit vectors moving on the surface of  the sphere. Chandra et al  \cite{chandra2019continuous} showed that for $N$ odd
the onset of synchronization happens through a first order phase transition. They also proposed an extension of the
Ott-Antonsen ansatz to the N-dimensional system \cite{chandra2019complexity} by modifying slightly the original function
proposed in \cite{Ott2008}. Here we take a different approach to construct an ansatz for the system: instead of guessing
the form of the distribution of oscillators on the sphere, we expand the distribution in spherical harmonics and make an
ansatz for the expansion coefficients, similar to the construction of the 2D case. We obtain a function that differs
from the one derived in \cite{chandra2019continuous} and leads to a simpler relation between the ansatz and the order
parameter. We derive a differential equation for the ansatz parameters and the phase transition diagram for three types
of distributions of natural frequencies. Our results agree well with numerical simulations using $N=10000$ oscillators.

This paper is organized as follows: in section II we re-derive the reduced equations for the 2D Kuramoto model using 
unit vectors. We show that Ott and Antonsen's ansatz describes the phase transition for  any distribution of frequencies, not
only the Lorentzian. In section III we briefly review the 3D Kuramoto model, propose our ansatz and derive the
differential equations for the ansatz parameters using the continuity equation. We then consider the case of identical 
oscillators and then the equilibrium solutions for general distributions of natural frequencies to study the behavior of the order parameter in terms of the coupling intensity, where we find a first order phase transition in dimension 3. In section IV we discuss
the differences and similarities between our reduced system and that proposed in \cite{chandra2019complexity}.

% %%%%%%%%%%%%%%%%%%%%%%%%%%%%%%%%%%%%%%%%%%%%%%%%%%%%
% %%%%%%%%%%%%%%%%%%%%%%%%%%%%%%%%%%%%%%%%%%%%%%%%%%%%
\section{2D Kuramoto model}

In this section we rewrite the original Kuramoto model in vector form and show how the Ott-Antonsen ansatz can be used to
characterize the behavior of the order parameter as a function of the coupling constant in equilibrium. 

\subsection{Vector formulation}

Following Chandra et al, we describe the oscillators by unit vectors $\vec{\sigma_i} = (\cos{\phi_i},\sin{\phi_i}).$
It is easy to show that if  $\phi_i$ satisfies Kuramoto's equation (\ref{kuramoto}) then
\begin{equation}
\frac{d \vec{\sigma_i}}{d t} = \mathbf{W}_i \vec{\sigma_i} + \frac{K}{N} \sum_j [\vec{\sigma_j} - (\vec{\sigma_i}\cdot \vec{\sigma_j}) \vec{\sigma_i}]
\end{equation}
where  $\mathbf{W}_i$ is an anti-symmetric matrix containing the natural frequency $\omega_i$:
\begin{equation}
\mathbf{W}_i = \left( 
\begin{array}{cc}
0 & -\omega_i \\
\omega_i & 0
\end{array}
\right).
\label{wmat}
\end{equation}
The complex order parameter  $z$, Eq.(\ref{paraord}), is replaced by the vector
\begin{equation}
\vec{p} = \frac{1}{N}\sum_i \vec{\sigma_i}
\label{vecpar}
\end{equation}
describing the center of mass of the system.

%ok
We also consider an extension of the model where the coupling constant $K$ is replaced  be a $2 \times 2$ matrix ${\mathbf K}_i $  with elements $K_{ij}=K_{ij}(\phi_i)$ that might depend on $\phi_i$. It is convenient to define an auxiliary vector $\vec{q}_i = {\mathbf K}_i \, \vec{p}$ and rewrite the generalization of the dynamical equations as
\begin{equation}
\frac{d \vec{\sigma_i}}{d t} = \mathbf{W}_i \vec{\sigma_i} +  [ \vec{q}_i - (\vec{\sigma_i}\cdot \vec{q}_i) \vec{\sigma_i}].
\label{kuramotogenk}
\end{equation}
Norm conservation, $|\vec{\sigma_i}|=1$, is guaranteed for any set of regular matrices ${\mathbf K}_i$ (as can be seen by taking the scalar product of Eq.(\ref{kuramotogenkw}) with $\vec{\sigma_i}$). Moreover,
the equations of motion are invariant under the transformation $\vec{q}_i \rightarrow \vec{q}_i + a_i \vec{\sigma_i}$, for any function  $a_i(\phi_i)$.  It is also convenient to add a third (artificial) dimension and write $\vec{\sigma_i} = (\cos{\phi_i},\sin{\phi_i,0})$  and $\vec{\omega}_i = (0,0,\omega_i).$ With these definitions we obtain
\begin{equation}
\frac{d \vec{\sigma_i}}{d t} = \vec\omega_i \times \vec{\sigma_i} +  [ \vec{q}_i - (\vec{\sigma_i}\cdot \vec{q}_i) \vec{\sigma_i}].
\label{kuramotogenkw}
\end{equation}
%
%ok
% %%%%%%%%%%%%%%%%%%%%%%%%%%%%%%%%%%%%%%%%%%%%%%%%%%%%
% %%%%%%%%%%%%%%%%%%%%%%%%%%%%%%%%%%%%%%%%%%%%%%%%%%%%
\subsection{Continuity equation}

In the limit of infinitely many oscillators we define $f(\omega,\phi,t)$ as the density of oscillators with natural frequency $\omega$ 
at position $\phi$ in time $t$. It satisfies 
\begin{equation}
\int_0^{2\pi} d \phi f(\omega,\phi,t) = g(\omega),
\end{equation}
where $g(\omega)$ is the distribution of natural frequencies and
\begin{equation}
\int_0^{2\pi} d \phi  \int d \omega f(\omega,\phi,t) = \int g(\omega) d \omega = 1.
\end{equation}

Conservation of oscillator number implies the continuity equation
\begin{equation}
\frac{\partial f}{\partial t} +  \frac{\partial (f v_\phi)}{\partial \phi}   = 0
\label{cont1}
\end{equation}
with velocity field
\begin{equation}
\vec{v} =  \omega \times \hat{r} +  \vec{q} - (\hat{r} \cdot \vec{q}) \hat{r}= (\omega + q_\phi) \hat{\phi}  \equiv v_\phi \hat{\phi}.
\end{equation}
Equation (\ref{cont1}) can be written explicitly as  
\begin{equation}
	\frac{\partial f}{\partial t} + (\omega+q_\phi) \frac{\partial f}{\partial \phi} - f q_r = 0
	\label{cont}
\end{equation}
where $q_r  = \vec{q} \cdot \hat{r}$, $q_\phi  = \vec{q} \cdot \hat{\phi}$, $\hat{r} = (\cos \phi, \sin \phi)$ and $\hat{\phi}=(-\sin \phi, \cos \phi)$  (we shall omit the artificial third
dimension whenever possible). We also note that $\frac{\partial q_\phi}{\partial \phi} = -q_r$. Eq.(\ref{vecpar}) for
the order parameter becomes 
\begin{equation}
\vec{p}(t) = \int \hat{r}(\phi) f(\omega,\phi,t) d \phi  \, d \omega.
\label{param}
\end{equation}

%ok
% %%%%%%%%%%%%%%%%%%%%%%%%%%%%%%%%%%%%%%%%%%%%%%%%%%%%
% %%%%%%%%%%%%%%%%%%%%%%%%%%%%%%%%%%%%%%%%%%%%%%%%%%%%
\subsection{Ansatz for density function}

The density of oscillators is a periodic function of $\phi$ and, therefore, can be expanded in Fourier series:
\begin{eqnarray}
f(\omega,\phi,t) = \frac{g(\omega)}{2\pi} \sum_{m=-\infty}^{\infty} a_m(\omega,t) e^{im\phi}.
\end{eqnarray}
Ott \& Antonsen \cite{Ott2008} showed that the ansatz $a_m = \rho^{|m|} e^{-im\Phi}$ is self-consistent, in the sense
that it preserves this form at all times, remaining in this restricted subset of density functions. The ansatz is,
therefore, parametrized by $\rho(\omega,t)$ and $\Phi(\omega,t)$: 
\begin{eqnarray}
f(\omega,\phi,t) = \frac{g(\omega)}{2\pi} \sum_{m=-\infty}^{\infty}  \rho^{|m|}  e^{im(\phi-\Phi)}.
\label{f1}
\end{eqnarray}
Summing the geometric series we obtain
\begin{eqnarray}
f(\omega,\phi,t) = \frac{g(\omega)}{2\pi} \frac{(1-\rho^2)}{1+\rho^2-2 \rho \cos \xi}
\label{f2}
\end{eqnarray}
where $\xi = \phi - \Phi$. It is convenient to define the vector $\vec{\rho} = \rho (\cos \Phi, \sin \Phi,0)$ so that $\cos \xi = \cos (\phi-\Phi) = \hat{r} \cdot \hat{\rho}$.

Substituting Eq.(\ref{f1}) into (\ref{param}) we see that only the terms  with $m=1$ and $m=-1$ contribute to the integral:
\begin{eqnarray}
\vec{p}(t) &=& \int \frac{g(\omega)}{2 \pi} \hat{r}(\phi) \rho \left(e^{i \xi} + e^{-i \xi}\right) d \phi  \, d \omega \nonumber 
   = \int \frac{g(\omega)}{\pi} \left[ \int_0^{2\pi} d \phi  \, \hat{r}(\phi)  \hat{r}^T(\phi) \right]  \vec{\rho} \, d \omega \nonumber 
\end{eqnarray}
where the dyadic product  $\hat{r}(\phi)  \hat{r}^T(\phi)$ is the projector in the $\hat{r}$ direction. 
It is easy to check that the integral between brackets results $\pi {\mathbf 1}$ and
\begin{equation}
\vec{p}(t) = \int \vec{\rho}(\omega,t)  g(\omega) \, d \omega.
\label{paramg}
\end{equation}

This is a key equation, that connects the parameters of the ansatz with the order parameter. The
specific relation depends on the form of $g(\omega)$. We will also use polar variables for the
order parameter, $\vec{p} = p (\cos \Psi, \sin \Psi,0)$ and $\nu = \phi - \Psi$.

%ok
% %%%%%%%%%%%%%%%%%%%%%%%%%%%%%%%%%%%%%%%%%%%%%%%%%%%%
% %%%%%%%%%%%%%%%%%%%%%%%%%%%%%%%%%%%%%%%%%%%%%%%%%%%%
\subsection{Equations of motion for ansatz parameters}

The derivatives of $f$ are
\begin{equation}
\frac{\partial f}{\partial \phi} = - \frac{\partial f}{\partial \Phi} = \frac{2 \rho_\phi (1-\rho^2 )}{ D^2},
\end{equation}
\begin{equation}
\frac{\partial f}{\partial \rho} = - \frac{2 [2\rho - (1+\rho^2) \cos \xi]}{ D^2},
\end{equation}
where we defined $D=1+\rho^2-2 \rho \cos \xi.$

We can write Eq.(\ref{cont}) explicitly as 
\begin{equation}
\frac{\partial f}{\partial \rho} \dot{\rho} + \frac{\partial f}{\partial \Phi} \dot{\Phi}  + (\omega+q_\phi) \frac{\partial f}{\partial \phi} - f q_r = 0
\label{contid1}
\end{equation}
Replacing the derivatives we find, after some simplifications presented in Appendix \ref{app1}, the
following equation for the dynamics of $\vec\rho$
\begin{eqnarray}
	\dot{\vec{\rho}}  =  \vec{\omega} \times \vec{\rho} + \frac{1}{2}(1+\rho^2) {\mathbf K} \vec{p} -  ({\mathbf K}  \vec{p} \cdot \vec{\rho})  \vec{\rho}.
	\label{eqmf}
\end{eqnarray}
Taking the scalar product with $\hat{\rho}$ we also get
\begin{eqnarray}
	\dot{\rho} &=& \frac{1}{2} (1-\rho^2)  ({\mathbf K} \vec{p} \cdot \hat{\rho}) . 
	\label{eqmff}
\end{eqnarray}
%

% %%%%%%%%%%%%%%%%%%%%%%%%%%%%%%%%%%%%%%%%%%%%%%%%%%%%
% %%%%%%%%%%%%%%%%%%%%%%%%%%%%%%%%%%%%%%%%%%%%%%%%%%%%
\subsection{Equilibrium solutions}

In this subsection we particularize for the case ${\mathbf K} = K {\mathbf 1} $. If $K=0$ the only equilibrium
solution is $\rho=p=0$. For $K \neq 0$ we can solve Eq.(\ref{eqmf}) with $	\dot{\vec{\rho}} = 0$ for a given $\vec{p}$
and check that the solution allows the determination of $\vec{p}$ via Eq.(\ref{paramg})  self-consistently. If $\rho
\neq 0$, then either $\rho=1$, and we only need to find the direction $\hat{\rho}$, or
$\vec{\rho}$ is perpendicular to $\vec{p}$ (see Eq.(\ref{eqmff})). Fixing $\hat{x}$ as the direction of $\vec{p}$, so that $\vec{p} = p
\hat{x}$ and $\vec{p} \cdot \hat{\rho} = p \cos \Phi$, and setting $\rho=1$ we find for the $\hat{x}$ and $\hat{y}$
components of Eq.(\ref{eqmf}) 
\begin{eqnarray}
	-\omega \sin\Phi + Kp -Kp \cos^2 \Phi = 0 \\
	\omega \cos \Phi -Kp \cos\Phi \sin \Phi =0.
\end{eqnarray}

\begin{figure*}
	\centering 
	\includegraphics[scale=0.4]{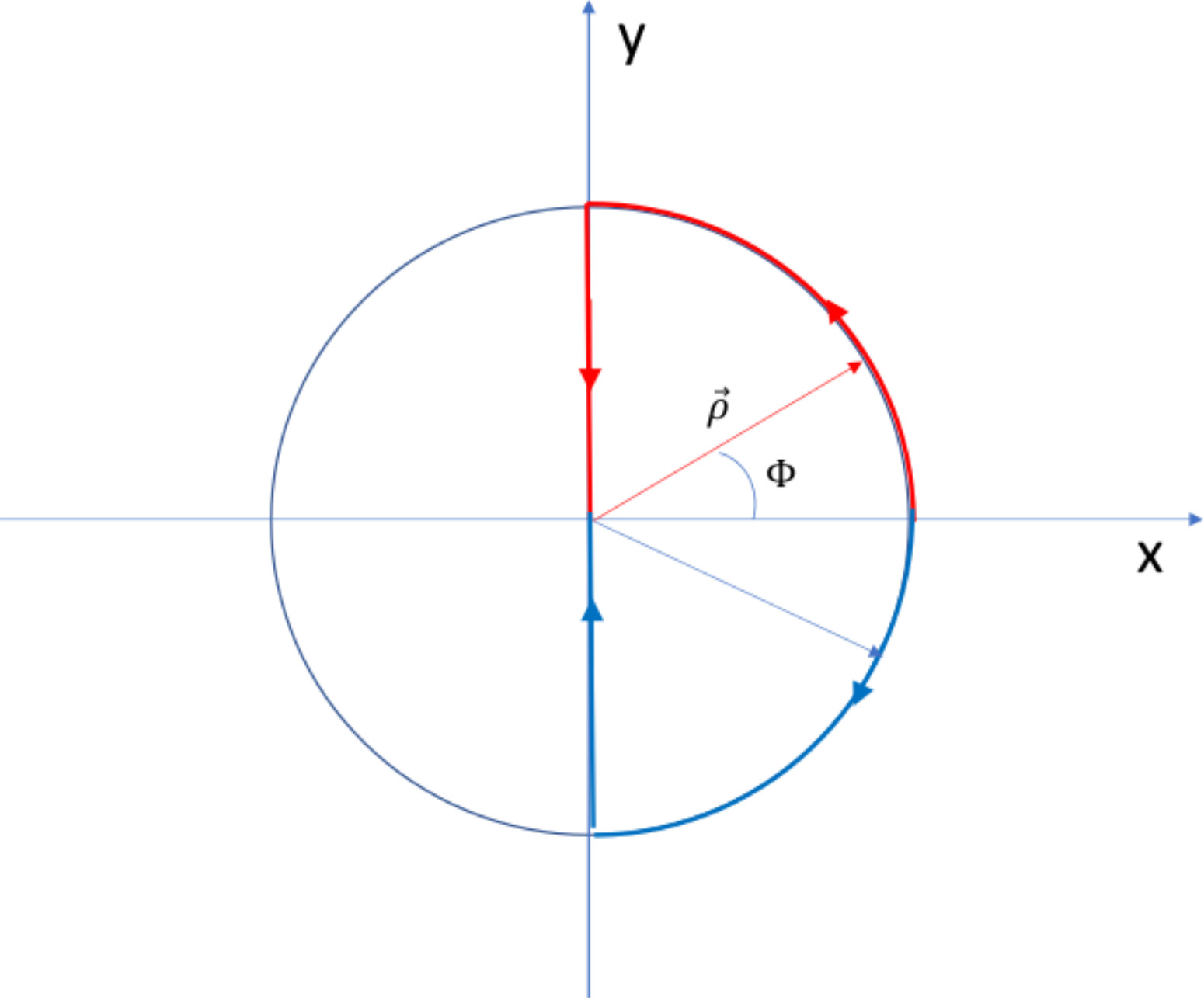} 
	\caption{Solution $\vec{\rho}(\omega)$ for $\omega >0$ (red). For $0 < \omega < Kp$ $\vec{\rho}$ stays on the unit circle, starting at 
	$\Phi=0$ and moving towards $\Phi=\pi/2$. For  $\omega > Kp$  $\vec{\rho}$ points in the y-direction 
	and decreases to zero as $\omega \rightarrow \infty$. For $\omega < 0$ (blue) the solution is reflected on the x-axis.} 
	\label{fig1}
\end{figure*}

The solution is $\sin \Phi = \frac{\omega}{Kp}$ if $|\omega| \leq Kp$. For $|\omega| > Kp$ we set $\vec{\rho}=\rho \hat{y}$ and get 
$\rho = \omega/Kp[1\pm \sqrt{1-K^2p^2/\omega^2}] $.
One can check that the stable solution is the one with the minus sign. The final solution, therefore is
\begin{equation}
	\vec{\rho} (\omega)= \frac{\omega}{Kp} \left\{
	\begin{array}{l}
		\sqrt{\frac{K^2p^2}{\omega^2} - 1} \, \hat{x}\,  +\,\hat{y} \qquad \mbox{for} \quad |\omega| \leq Kp \\ \\
		  \left[1-\sqrt{1-\frac{K^2p^2}{\omega^2}}\right] \hat{y} \qquad \mbox{for} \quad |\omega| > Kp .\\
	\end{array}
	\right.
\end{equation}
Figure \ref{fig1} illustrates the dependence of $\vec{\rho}$ on $\omega$.

The order parameter can now be computed for any distribution function $g(\omega)$. Assuming $g(\omega)$ to be symmetric around $\omega=0$ we obtain
\begin{eqnarray}
	p = Kp \int_{-\pi/2}^{\pi/2} \cos^2(\Phi) g(Kp\sin\Phi) d \Phi
	\label{orderp2}
\end{eqnarray}
which is the result found by Kuramoto. Notice that this was obtained from the ansatz, not from the exact equations.
The trivial solution is $p=0$. The non-trivial solution is obtained canceling $p$
on each side and solving for $p$. The critical value of $K$, where the non-trivial solution starts, is obtained by setting
$p=0$ in the resulting equation, leading to $K_c = 2/\pi g(0)$. Expanding $g$ around $p=0$ to second order we find
\begin{equation}
p= \sqrt{\frac{16}{\pi K^3 g''(0)}(1-K/K_c)}, 
\end{equation}
which is exactly what was found by Kuramoto. 

\begin{figure*}
	\centering 
	\includegraphics[width=0.45\linewidth]{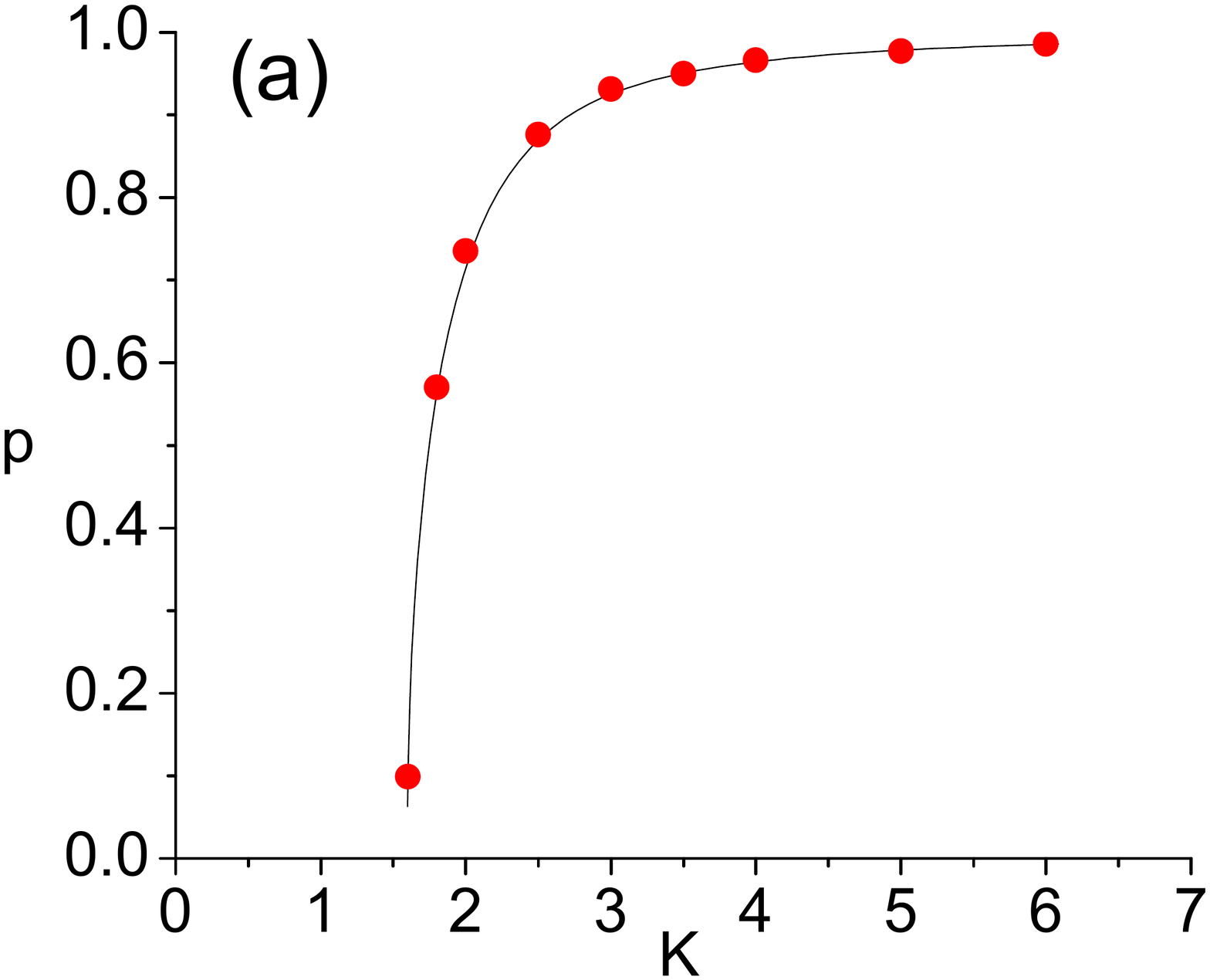} 
	\includegraphics[width=0.45\linewidth]{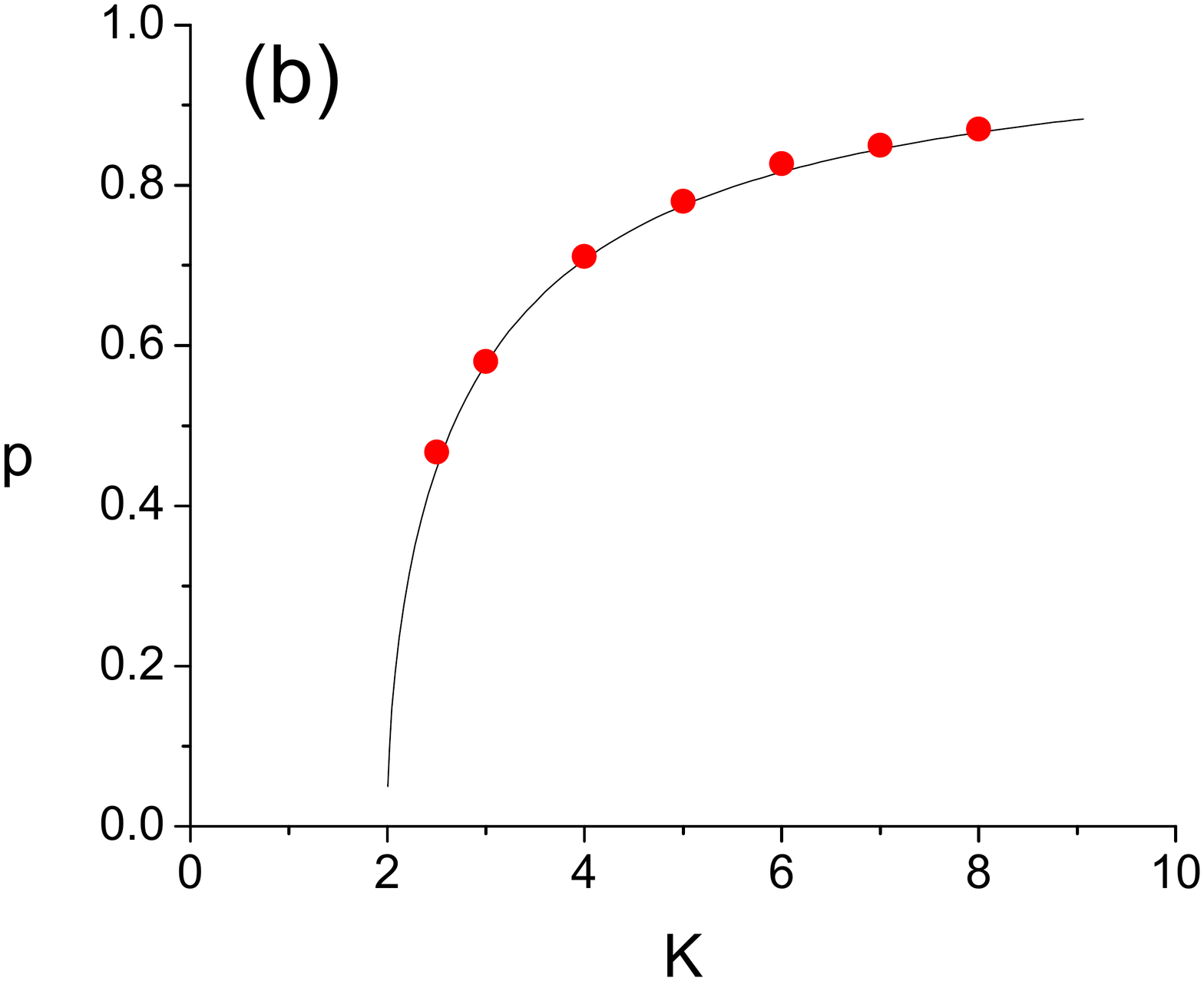} 
	\caption{Order parameter as a function of the coupling constant for the Gaussian (a) and Lorentzian (b) distribution of natural frequencies. 
	Lines show the analytical result and points the result of simulations with $N=1000$ oscillators.} 
	\label{fig2}
\end{figure*}

The complete $p \times K$ phase diagram can be constructed for specific distributions $g(\omega)$ \cite{hu2014exact}. Noting that the
right hand side of Eq.(\ref{orderp2}) is a function $F(Kp)$ the parametric plot $p=p(K)$ can be
obtained as $(K,p) = (x/F(x),F(x))$ for $x \in (0,\infty)$. For Gaussian distribution with unit variance we get
\begin{equation}
	F(x) = \sqrt{\frac{\pi}{8}} \, x e^{-x^2/4} [I_0(x^2/4)+ I_1(x^2/4)]
\end{equation}
where $I_n(x)$ is the modified Bessel function of the first kind. For the Lorentzian distribution with unit width we get
\begin{equation}
	F(x) = \frac{1}{x}\left[\sqrt{1+x^2} - 1\right].
\end{equation}
In both cases $K_c$ can be obtained as $\lim_{x \rightarrow 0} x/F(x)$ which results $K_c=\sqrt{8/\pi}$ and $K_c=2$ for
the Gaussian and Lorentzian distributions respectively. Figure \ref{fig2} compares these results with numerical simulations. We
note that numerically solving the pair of Eqs.(\ref{kuramotogenkw}) is much faster than the original Eq.(\ref{kuramoto}).

% %%%%%%%%%%%%%%%%%%%%%%%%%%%%%%%%%%%%%%%%%%%%%%%%%%%%
% %%%%%%%%%%%%%%%%%%%%%%%%%%%%%%%%%%%%%%%%%%%%%%%%%%%%
\section{3D Kuramoto model}

The 3D model is a direct extension of the vector 2D model and represents unit vectors $\vec{\sigma_i}$ rotating 
on the surface of a sphere. The equations are \cite{chandra2019continuous}
\begin{equation}
	\frac{d \vec{\sigma_i}}{d t} = \mathbf{W}_i \vec{\sigma_i} + \frac{1}{N}  
	\sum_j [\mathbf{K}_i \vec{\sigma_j} - (\vec{\sigma_i}\cdot \mathbf{K}_i \vec{\sigma_j}) \vec{\sigma_i}]
	\label{3dkura}
\end{equation}
where  $\mathbf{W}_i$ is a $3 \times 3$ anti-symmetric matrix 
\begin{equation}
	\mathbf{W}_i = \left( 
	\begin{array}{ccc}
		0 & -\omega_{3i}  &  \omega_{2i} \\
		\omega_{3i} & 0  &  -\omega_{1i} \\
		-\omega_{2i} & \omega_{1i} & 0 
	\end{array}
	\right)
	\label{wmat3}
\end{equation}
and $\mathbf{K}_i$ is the $3 \times 3$ coupling matrix. There are now three independent frequencies
for each oscillator (or unit vector) that are taken from a normalized distribution
$G(\vec{\omega})$.  The order parameter is defined as in the 2D case as the system's  center of mass 
\begin{equation}
	\vec{p} = \frac{1}{N}\sum_i \vec{\sigma_i}.
	\label{vecpar3}
\end{equation}
Defining the 3D vector $\vec w_i =(\omega_{1i},	\omega_{2i}, \omega_{3i})$, renaming  $	\vec\sigma \equiv\hat r$ 
and defining
\begin{eqnarray}
	{\mathbf K}_ i \, \vec p \equiv \vec q_i 
\end{eqnarray}
we obtain the dynamical equations
\begin{eqnarray}
	\dot{\hat{r}}_i = \vec w_i \times \hat{r}_i+ \vec q_i -(\hat{r}_i\cdot \vec q_i)\hat{r}_i.
	\label{kura3}
\end{eqnarray}
As in the 2D case, the equations of motion are invariant under the transformation $\vec q_i
\rightarrow \vec q_i + a_i \hat r_i $ and we shall use this freedom later.

% ok

% %%%%%%%%%%%%%%%%%%%%%%%%%%%%%%%%%%%%%%%%%%%%%%%%%%%%
% %%%%%%%%%%%%%%%%%%%%%%%%%%%%%%%%%%%%%%%%%%%%%%%%%%%%
\subsection{Continuity equation}

The continuity equation on the sphere is
\begin{equation}
	\frac{\partial f}{\partial t} + \frac{1}{\sin\theta} \frac{\partial}{\partial \theta}
	(f\sin\theta v_\theta)+ \frac{1}{\sin\theta}\frac{\partial}{\partial\phi}(f v_\phi) = 0
\end{equation}
with velocity field given by Eq. (\ref{kura3}):
\begin{eqnarray}
	\vec v = \vec{\omega} \times \hat{r} + \vec q -(\hat{r}\cdot \vec q)\hat{r}.
\end{eqnarray}
Computing the partial derivatives the continuity equation takes the form
\begin{eqnarray}
\frac{\partial f}{\partial t} -2fq_r + (\omega_\phi+q_\theta)\frac{\partial f}{\partial\theta}+ \frac{1}{\sin\theta}(-\omega_\theta+q_\phi)\frac{\partial f}{\partial\phi} = 0
	\label{fv1}
\end{eqnarray}
and the order parameter becomes:
\begin{eqnarray}
	\vec p(t) = \int\hat r(\theta',\phi') f(\theta',\phi',\vec\omega, t) d\Omega' d^3\omega 
\end{eqnarray}

%ok

\subsection{Ansatz for the density function}

In 3D we can expand the density function in spherical harmonics:
\begin{eqnarray}
	f(\vec\omega,\theta,\phi,t) = G(\vec\omega) \sum_{l=0}^{\infty}\sum_{m=-l}^{l} f_{lm} Y_{lm}(\theta,\phi).
\end{eqnarray}
Now we propose an ansatz analogous to the 2D case, but with three parameters $\rho$, $\Theta$, $\Phi$, with 
the form $f_{lm} = \rho^l Y^{*}_{lm}(\Theta, \Phi)$. The parameters define a vector  $ \vec\rho(\vec\omega,t) = \rho (\sin\Theta\cos\Phi,\sin\Theta\sin\Phi, \cos\Theta)$. We write
\begin{eqnarray}
	f(\vec\omega,\theta,\phi,t) = \frac{G(\vec\omega)}{4\pi}\left[1+ 4\pi \sum_{l=1}^{\infty}\sum_{m=-l}^{l} \rho^l Y^{*}_{lm}(\Theta, \Phi) Y_{lm}(\theta,\phi)\right].
	\label{SphericalHarmonicsExpansion}
\end{eqnarray}
Note that for $\rho =1$ we obtain $f(\omega,\theta,\phi,t) = G(\vec\omega)
\frac{\delta(\theta-\Theta)\delta(\phi-\Phi)}{\sin \theta} $ indicating  full synchrony. For $\rho =0$, on the other
hand, $f(\omega,\theta,\phi,t) = \frac{G(\vec\omega)}{4\pi}$ and the oscillators are uniformly spread over the sphere.

To sum the series and calculate the distribution $f$ explicitly we use the relations 
\begin{eqnarray}
	\frac{4\pi}{2l+1}\sum_{m=-l}^{l}Y_{lm}(\hat y)Y^*_{lm}(\hat x) = P_l(\hat x \cdot\hat y) \\
	\sum_{l=0}^\infty y^lP_l(x) = \frac{1}{\sqrt{1+y^2-2xy}} 
\end{eqnarray}
Applying these relations to (\ref{SphericalHarmonicsExpansion}) we find
\begin{equation}
		f(\vec\omega,\theta,\phi,t) = \frac{G(\vec\omega)(1-\rho^2)}{4\pi(1+\rho^2-2\rho\hat r\cdot\hat\rho)^{{\frac{3}{2}}}}.
		\label{anssol}
\end{equation}
Notice that $f$ depends on $\Theta$ and $\Phi$ only through
\begin{eqnarray}
	 \hat r\cdot \hat\rho \equiv \cos\xi = \sin\theta\sin\Theta\cos(\phi-\Phi) + \cos\theta\cos\Theta.
\label{ftp}
\end{eqnarray}
In addition, using the relation 
\begin{eqnarray}
	\frac{4\pi}{3}\sum_{m=-1}^{1}Y_{1m}(\hat y)Y^*_{1m}(\hat x)= \hat x \cdot\hat y
\end{eqnarray}
and the orthogonality of the spherical harmonics we can write the order parameter as
\begin{equation}
		\vec p =\int G(\vec\omega)\rho \frac{3}{4\pi}\left[\int \hat r'\hat r'^T d\Omega'\right]\cdot\hat \rho d^3\omega    
\end{equation}
The angular integral of the dyadic matrix can be easily performed and results $ (4\pi/3)\mathbf{1}$. We obtain
\begin{eqnarray}
	\vec p=\int G(\vec\omega) \, \vec\rho \, d^3\omega.
	\label{eqp3d}
\end{eqnarray}
This is formally identical to the 2D case (see Eq. (\ref{paramg})), and much simpler than the relation found in
\cite{chandra2019complexity}. Note that this simplification was only possible because we applied orthogonality properties of the spherical harmonics. Also, Eq.(\ref{anssol}) should be contrasted with the expression derived in
\cite{chandra2019complexity}, where the denominator is raised to the power 2, instead of $3/2$, and the numerator
$(1-\rho^2)$ to power 2, instead of 1. These differences will have consequences for the derivation of the equations of
motion.

%ok

% %%%%%%%%%%%%%%%%%%%%%%%%%%%%%%%%%%%%%%%%%%%%%%%%%%%%
% %%%%%%%%%%%%%%%%%%%%%%%%%%%%%%%%%%%%%%%%%%%%%%%%%%%%
\subsection{Continuity equation for ansatz distribution}

The partial derivatives of $f$ are
\begin{eqnarray}
	\frac{\partial f}{\partial\rho}=\frac{\cos\xi(3+\rho^2)-\rho(5-\rho^2)}{D^{\frac{5}{2}}}
	\label{der3}
\end{eqnarray}
and 
\begin{eqnarray}
	\frac{\partial f}{\partial x}=\frac{\partial f}{\partial\cos\xi}\frac{\partial\cos\xi}{\partial x} = \frac{3\rho(1-\rho^2)}{D^{\frac{5}{2}}}\frac{\partial\cos\xi}{\partial x},
	\label{der2}
\end{eqnarray}
for $x=\theta, \phi, \Theta, \Phi$ where $D \equiv 1+\rho^2-2\rho\cos\xi$ and
\begin{equation}
	\begin{split}
		\frac{\partial\cos\xi}{\partial\theta}&= \frac{\rho_\theta}{\rho}\qquad\qquad\quad\frac{\partial\cos\xi}{\partial\phi} = \sin\theta\frac{\rho_\phi}{\rho}
		\\
		\qquad\frac{\partial\cos\xi}{\partial\Theta} &= \hat{r}\cdot \hat{\Theta}\qquad\qquad
		\frac{\partial\cos\xi}{\partial\Phi} = \sin \Theta \hat{r} \cdot \hat{\Phi} 
	\end{split}
	\label{der1}
\end{equation}

Writing  the time derivative of $f$ as
\begin{eqnarray}
	\frac{\del f}{\del t}=\frac{\del f}{\del\rho}\dot\rho+ \frac{\del f}{\del\Theta}\dot\Theta+ \frac{\del f}{\del\Phi}\dot\Phi
	\label{timederiv}
\end{eqnarray}
we obtain
\begin{equation}
	\begin{split}
		D^{\frac{5}{2}}\frac{\del f}{\del t}=\left[\cos\xi(3+\rho^2)-\rho(5-\rho^2)\right]\dot\rho 
		+ 3\rho(1-\rho^2) \hat{r} \cdot [\dot{\Theta} \hat{\Theta}+ \sin \Theta \dot{\Phi}\hat{\Phi}] .
	\end{split}
\end{equation}

Using eqs.(\ref{fv1}) and (\ref{der3})-(\ref{der1}) we can also write (see Appendix \ref{app2})
\begin{equation}
	\begin{split}
		D^{\frac{5}{2}}  \nabla(f\vec v) &= 3\rho(1-\rho^2)\hat{r} \cdot (\hat{\rho} \times \vec{\omega}) + (1-\rho^2)\left[-2 q_r(1+\rho^2) + 4 q_r \rho_r + 3 q_\theta \rho_\theta + 3 q_\phi \rho_\phi\right].
		\label{fv4}
	\end{split}
\end{equation}

\subsection{Compatibility conditions}

At this point we need to specify the vector $\vec{q}$ that will describe the coupling between the oscillators. The natural choice  $\vec{q} = {\mathbf K} \vec{p}$, does not work. To see this consider, for instance ${\mathbf K}=K {\mathbf I}$ or $\vec{q} = K \vec{p}$. In this case the last three terms in Eq.(\ref{fv4}) become
\begin{equation}
	4\rho_r p_r + 3\rho_\theta p_\theta + 3\rho_\phi p_\phi = 3\vec\rho \cdot \vec{p} + \rho_r p_r = 
	3\vec\rho \cdot \vec{p} + \rho p \cos\xi \cos \nu
\end{equation}
and the term $\cos\xi \cos \nu$ has no counterpart in the continuity equation. Chandra et al
\cite{chandra2019complexity} have a similar problem, that they solve by choosing the exponent of their tentative ansatz
function. Here we do not have this freedom. However, there is still a way out, using the invariance of the exact
equations under changes in the radial part of $\vec{q}$. We can cancel the unwanted terms without altering the exact
equations of motion if we choose 
\begin{equation}
	\vec q =  \left[\mathbf{1} - \frac{1}{4} \hat{r} \hat{r}^\top \right] {\mathbf K} \vec p + \beta \hat{r}
		\label{gauge}
\end{equation}
where $\hat{r} \hat{r}^\top$ is the projection in the radial direction and the coefficient $\beta$ will be chosen later to ensure consistency of the equations of motion. With this choice
Eq.(\ref{fv4}) becomes 
\begin{equation}
	\begin{split}
		D^{\frac{5}{2}}  \nabla(f\vec v) &= 	3\rho(1-\rho^2)  \hat{r} \cdot (\hat{\rho} \times \vec{\omega}) \\ 
		&+ 	(1-\rho^2)\left[-2 \beta (1+\rho^2) + 3 \vec{\rho} \cdot ({\mathbf K} \vec{p}) - \frac{3}{2} (1+\rho^2) \hat{r} \cdot  ({\mathbf K} \vec{p}) + 4 \beta \hat{r} \cdot \vec{\rho} \right].
	\end{split}
	\label{fv3}
\end{equation}

%ok 

%%%%%%%%%%%%%%%%%%%%%%%%%%%%%%%%%%%%%%%%%%%%%
\subsection{Equations of motion}    

The continuity equation for the ansatz function, after multiplying by $D^{5/2}$, is given by
\begin{equation}
	\begin{split}
		\left[\cos\xi(3+\rho^2)-\rho(5-\rho^2)\right]\dot\rho 
		+ 3\rho(1-\rho^2) \hat{r} \cdot [\dot{\Theta} \hat{\Theta}+ \sin \Theta \dot{\Phi}\hat{\Phi}]  
		+  3\rho(1-\rho^2) 	\hat{r} \cdot (\hat{\rho} \times \vec{\omega}) \\
		- (1-\rho^2)\left[-2 \beta (1+\rho^2) + 3 \vec{\rho} ({\mathbf K} \vec{p}) - \frac{3}{2} (1+\rho^2) \hat{r} \cdot  ({\mathbf K} \vec{p}) + 4 \beta \hat{r} \cdot \vec{\rho} \right]=0
	\end{split}
	\label{eqmot}
\end{equation}    
We simplify this equation in Appendix \ref{app3}. This  leads us to choose
\begin{equation}
	\beta = \frac{\vec{\rho} \cdot ({\mathbf K} \vec{p}) }{4}.
	\label{beta}
\end{equation}
and the final equations of motion become
\begin{equation}
	\begin{split}
		\dot{\vec{\rho}} = \vec{\omega} \times \vec{\rho}  + \frac{1}{2} (1+\rho^2) ({\mathbf K} \vec{p})  -  [\vec{\rho} \cdot ({\mathbf K} \vec{p}) ] \vec{\rho} .
	\end{split}
	\label{eqm1f}
\end{equation}   
From this equation it also follows that
\begin{equation}
	\begin{split}
		\dot{\rho}  = \frac{1}{2}(1-\rho^2)  [\hat{\rho} \cdot ({\mathbf K} \vec{p}) ].
	\end{split}
	\label{eqm2f}
\end{equation}    
The connection between $\vec{p}$ and $\vec{\rho}$ is given by Eq.(\ref{eqp3d}). Eq.(\ref{eqm1f}) is identical to the
equation obtained in \cite{chandra2019complexity}, but Eq.(\ref{eqp3d}) is not. In our formulation the order parameter
$\vec{p}(t)$ is the average over the distribution of natural frequencies of the ansatz parameter $\vec{\rho}(\omega,t)$,
just like the 2-dimensional case, whereas the expression in \cite{chandra2019complexity} includes an extra integral
that depends on the dimension of system. For $D=2$ both formulations recover the original result in \cite{Ott2008}.

%ok

% %%%%%%%%%%%%%%%%%%%%%%%%%%%%%%%%%%%%%%%%%%%%%%%%%%%%
% %%%%%%%%%%%%%%%%%%%%%%%%%%%%%%%%%%%%%%%%%%%%%%%%%%%%
\subsection{Identical oscillators with symmetric coupling and external forces}

\begin{figure*}
	\centering 
	\includegraphics[width=0.45\linewidth]{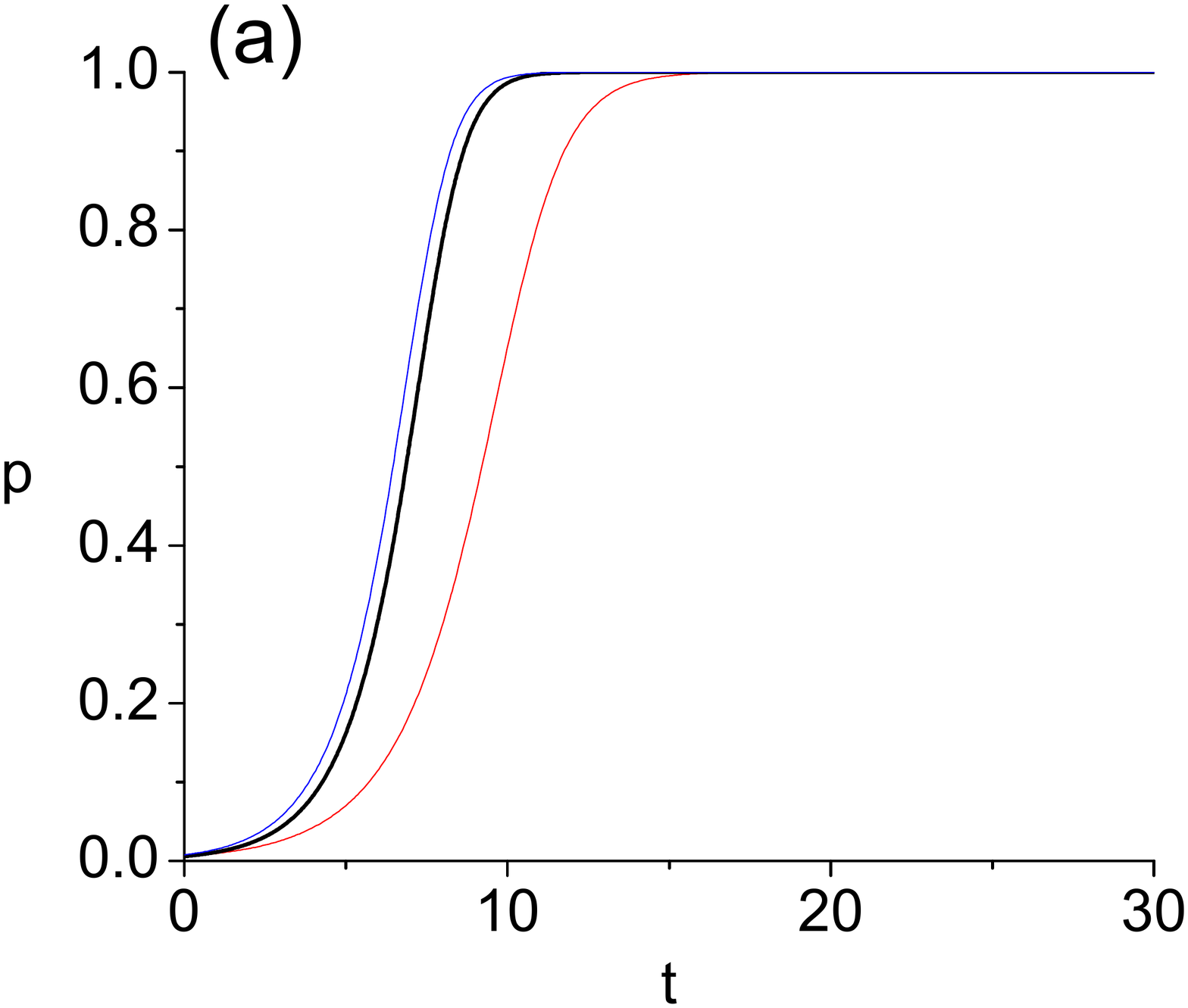} 
	\includegraphics[width=0.45\linewidth]{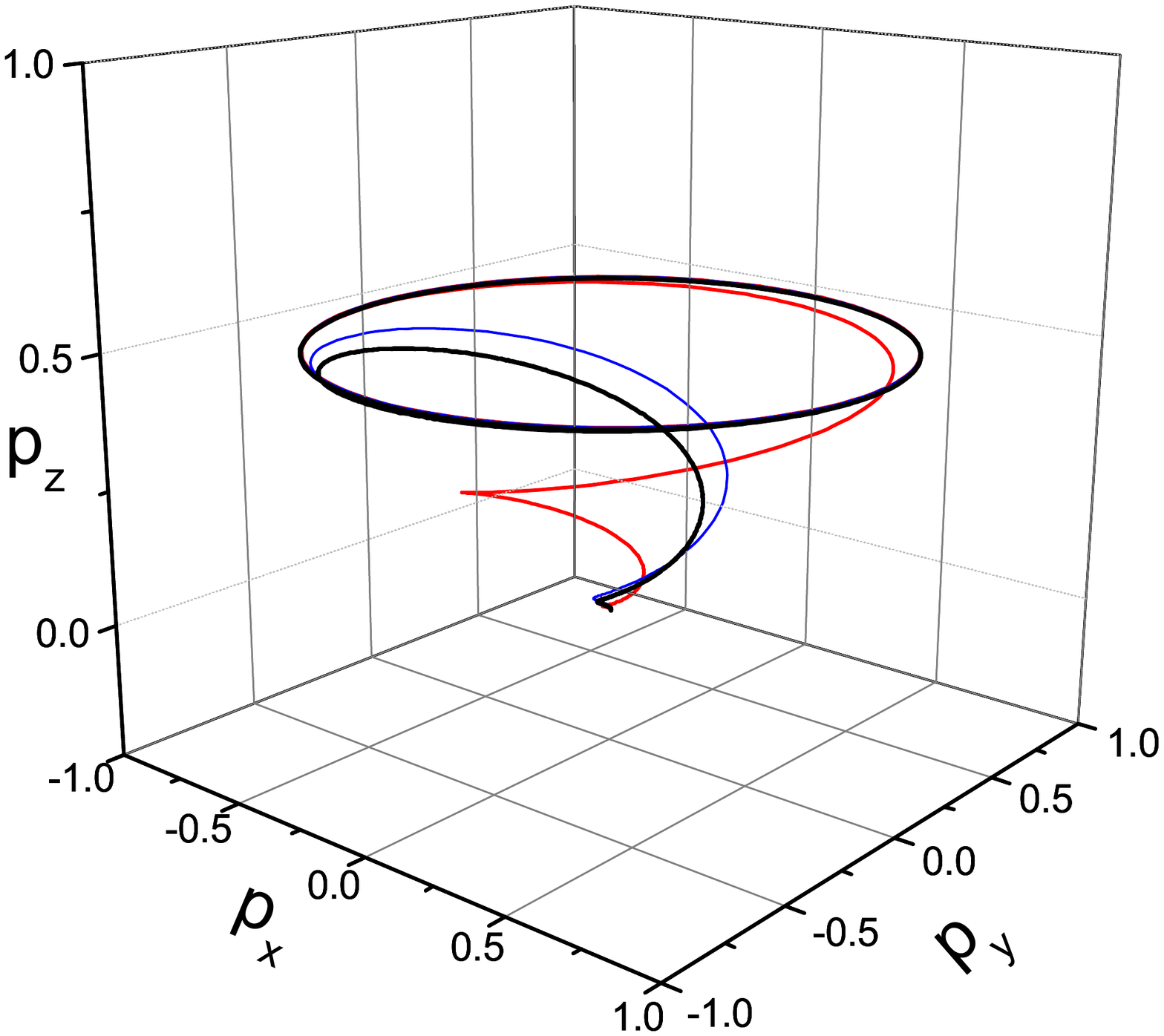} 
	\caption{Time evolution of the order parameter for identical oscillators with $\vec{\omega}=\hat{z}$ and $K=1$. (a) module
	$p(t)$; (b) evolution on the $p_x-p_y-p_z$ space. Thick black line shows the result of simulations with $N=10000$ oscillators.
	Thin red and blue line shows the ansatz solution using our approach and that from \cite{chandra2019complexity}, respectively.} 
	\label{fig3}
\end{figure*}

If all oscillators have identical natural frequencies, $G(\vec{\omega}) = \delta(\vec{\omega}-\vec{\omega}_0)$, then
$\vec{p}(t) = \vec{\rho}(\vec{\omega}_0,t)$. Moreover, if the coupling matrix is diagonal, $\mathbf{K} = K \mathbf{1}$, the
order parameter satisfies 
\begin{equation}
	\begin{split}
		\dot{\vec{p}} = \vec{\omega}_0 \times \vec{p}  + \frac{K}{2} (1-p^2)  \vec{p} \quad \quad \textrm{and} \quad\quad \dot{p}  = \frac{K}{2}(1-p^2) p.
	\end{split}
\end{equation}   
The equilibrium points are $p=0$, which is stable for $K<0$ and $p=1$, stable for $K>0$. Figure \ref{fig3} shows the
behavior of the order parameter comparing  the exact solution, obtained by a simulation with 10000 oscillators, the
solution based on our ansatz and the solution for the ansatz proposed in \cite{chandra2019complexity}, which we call Chandra's solution for
short.  In this case it is clear that our solution is delayed with respect to the exact (numerical) integration, whereas Chandra's
solution is only slightly advanced. Both reach the same equilibrium $p=1$.

Time dependent external forces can be included making ${\mathbf K} \cdot \vec{p}  \rightarrow {\mathbf K} \cdot \vec{p} 
+ \vec{F}$ in Eqs. (\ref{gauge}) and (\ref{beta})-(\ref{eqm2f}). For the case of diagonal coupling matrix this gives
\begin{equation}
	\begin{split}
		\dot{\vec{p}} = \vec{\omega}_0 \times \vec{p}  + \left[ \frac{K}{2} (1-p^2) - \vec{p}\cdot \vec{F} \right] \vec{p}
		+ \frac{1}{2}(1+p^2) \vec{F}
	\end{split}
\end{equation}   
and
\begin{equation}
	\begin{split}
		\dot{p}  =  \frac{1}{2} (1-p^2)(Kp + \vec{F}\cdot \hat{p}).
	\end{split}
\end{equation}    
%

%ok

% %%%%%%%%%%%%%%%%%%%%%%%%%%%%%%%%%%%%%%%%%%%%%%%%%%%%
% %%%%%%%%%%%%%%%%%%%%%%%%%%%%%%%%%%%%%%%%%%%%%%%%%%%%
\subsection{Equilibrium solutions for symmetric coupling}

Here we consider equilibrium solutions for $\mathbf{K} = K \mathbf{1}$ but general distributions of natural frequencies.
In this case Eqs. (\ref{eqm1f})  and (\ref{eqm2f}) simplify to:
\begin{equation}
	\begin{split}
		\dot{\vec{\rho}} = \vec{\omega} \times \vec{\rho}  + \frac{K}{2}[ (1+\rho^2)  \vec{p}  - 2 (\vec{\rho} \cdot \vec{p} ) \vec{\rho}] \quad \quad \textrm{and} \quad\quad \dot{\rho}  
		  = \frac{K}{2}(1-\rho^2)  (\hat{\rho} \cdot  \vec{p} ).
	\end{split}
	\label{eqm1f1}
\end{equation}   
We first consider the limit $K \rightarrow0$, or  $K=0^+$. According to Eqs. (\ref{eqm1f1}),  for $\vec{p} \neq 0$ equilibrium requires
$\rho=1$ and $\hat{\rho}$ parallel to $\hat{\omega}$, i.e., $\vec{\rho} = \pm \hat{\omega}$. However, only one of these
solutions is stable. To find out which we make  $\vec{\rho} = \pm \hat{\omega} + \delta \vec{\rho}$, or  $\rho = 1 +
\delta \rho$ and expand the module Eq.(\ref{eqm1f1}) to first order to obtain 
\begin{equation}
	\delta \dot{\rho} = -\frac{K}{2} \delta \rho (\vec{p} \cdot \hat{\rho}).
\end{equation}
Therefore, for each $\vec{\omega}$ the stable solution is the one for which $\vec{p} \cdot \hat{\rho} > 0$, i.e., the one
in the hemisphere defined by the direction of $\vec{p}$. Setting $\vec{p} = p \hat{z}$ we have for $G(\vec\omega) = g(\omega)/4 \pi$
\begin{eqnarray}
	p=\int G(\vec\omega) \rho_z d^3\omega = \frac{1}{4 \pi }\int_0^{2\pi} d\Phi \int_0^{\pi/2} 2 \sin  \Theta \cos \Theta d \Theta = \frac{1}{2},
	\label{rhoz}
\end{eqnarray}
where the factor 2 comes from the fact that, in the upper hemisphere, the stable solution is $\hat{\rho} = \hat{\omega}$
and in the lower hemisphere it is $\hat{\rho} = -\hat{\omega}$. So when we cross from one hemisphere to another both
$\cos(\theta)$ and $\hat{\rho}$ changes signal, making the integral over $0<\theta<\pi$ symmetrical with respect to
$\pi/2$.

For $K\neq 0$ we define the orthonormal set of vectors
\begin{eqnarray}
	\hat{\xi}_1 &=& (\cos \theta \cos \phi, \cos \theta \sin \phi, -\sin \theta)  \nonumber \\
	\hat{\xi}_2 &=& (-\sin \phi, \cos \phi, 0) \\
	\hat{\omega} &=& (\sin \theta \cos \phi, \sin \theta \sin \phi, \cos \theta) \nonumber 
\label{omegacoord}
\end{eqnarray}
and expand $\vec{\rho} = \alpha \hat{\xi}_1 + \beta \hat{\xi}_2 + \gamma \hat{\omega}.$
For $K=0$ we use $\alpha=\beta=0$ and $\gamma=1$, with $0 < \theta < \pi/2$. We again set $\vec{p} = p \hat{z}$ and 
obtain
\begin{equation}
	\vec{p} = -p \sin \theta \, \hat{\xi}_1 + p \cos \theta \, \hat{\omega}.
\end{equation}
Projecting Eq.(\ref{eqm1f}) with $	\dot{\vec{\rho}} =0$ in these directions we obtain
\begin{eqnarray}
	0 &=& -\beta - \bar{K} \sin \theta (1+ \rho^2)/2 - \bar{K} \alpha (- \alpha \sin \theta + \gamma \cos \theta) \nonumber \\
	0 & = &  \alpha - \bar{K}  \beta (- \alpha \sin \theta + \gamma \cos \theta)  	\label{abc1} \\
	0 &=&  \bar{K} \cos \theta(1+ \rho^2)/2 - \bar{K} \gamma (- \alpha  \sin \theta + \gamma \cos \theta). \nonumber
\end{eqnarray}
where $\rho^2=\alpha^2 + \beta^2 + \gamma^2$ and $\bar{K} = K p /\omega$.  These equations can be solved analytically: 
\begin{eqnarray}
	\beta &=& - \frac{(1+\bar{K}^2)}{2\bar{K} \sin\theta}\left[1-\sqrt{1-\frac{4\bar{K}^2 \sin^2 \theta}{(1+\bar{K}^2)^2}}\right] \\
	\alpha &=& -\sqrt{1+ \beta \left( \frac{1+ \bar{K}^2 \cos^2 \theta}{\bar{K} \sin\theta}\right)} \\ 
	\gamma &=& \frac{\bar{K} \beta}{\alpha} \cos\theta
	\label{anasol}
\end{eqnarray}
and satisfy $\rho^2=\alpha^2 + \beta^2 + \gamma^2 = 1$.

\begin{figure*}
	\centering 
	\includegraphics[width=0.32\linewidth]{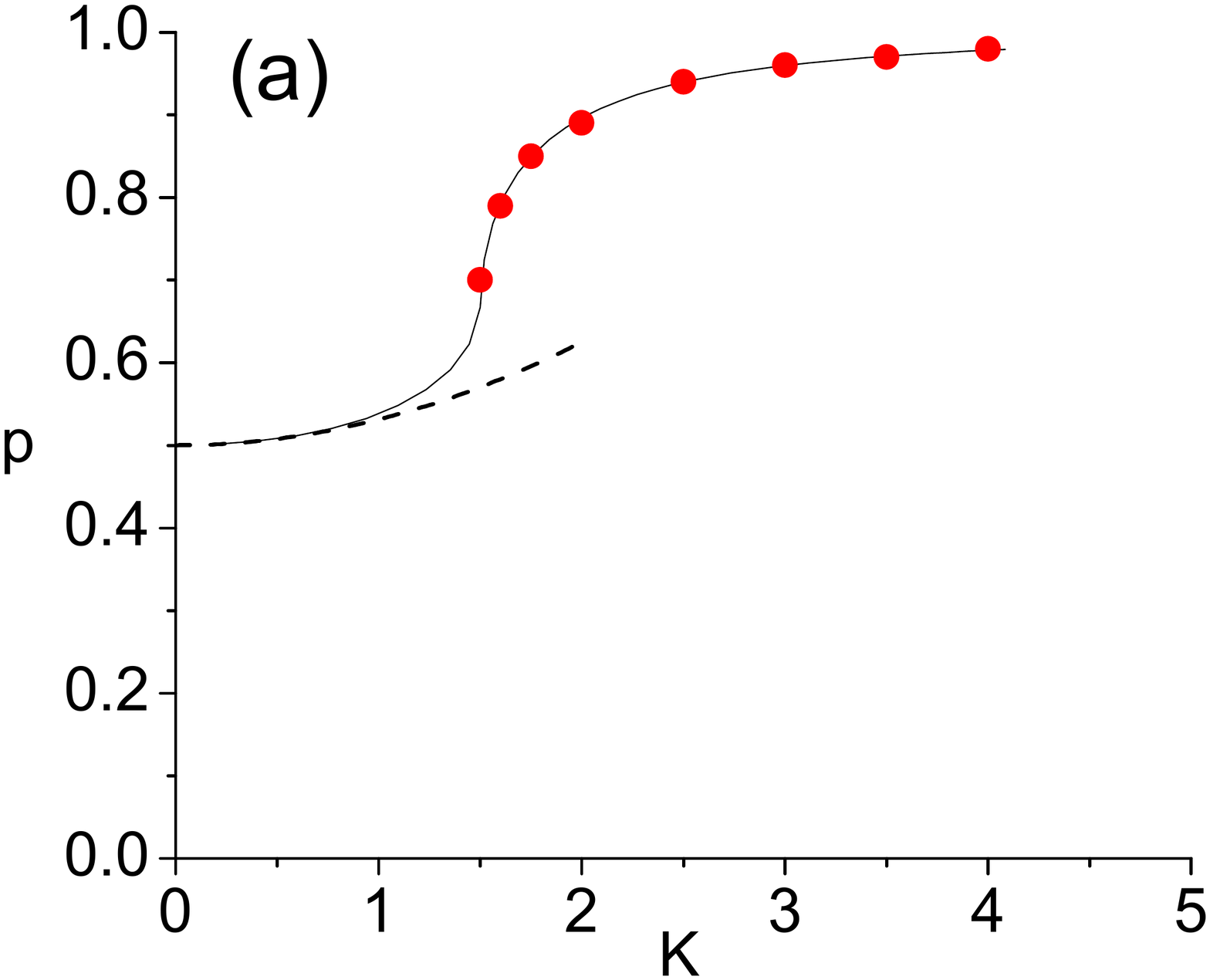} 
	\includegraphics[width=0.32\linewidth]{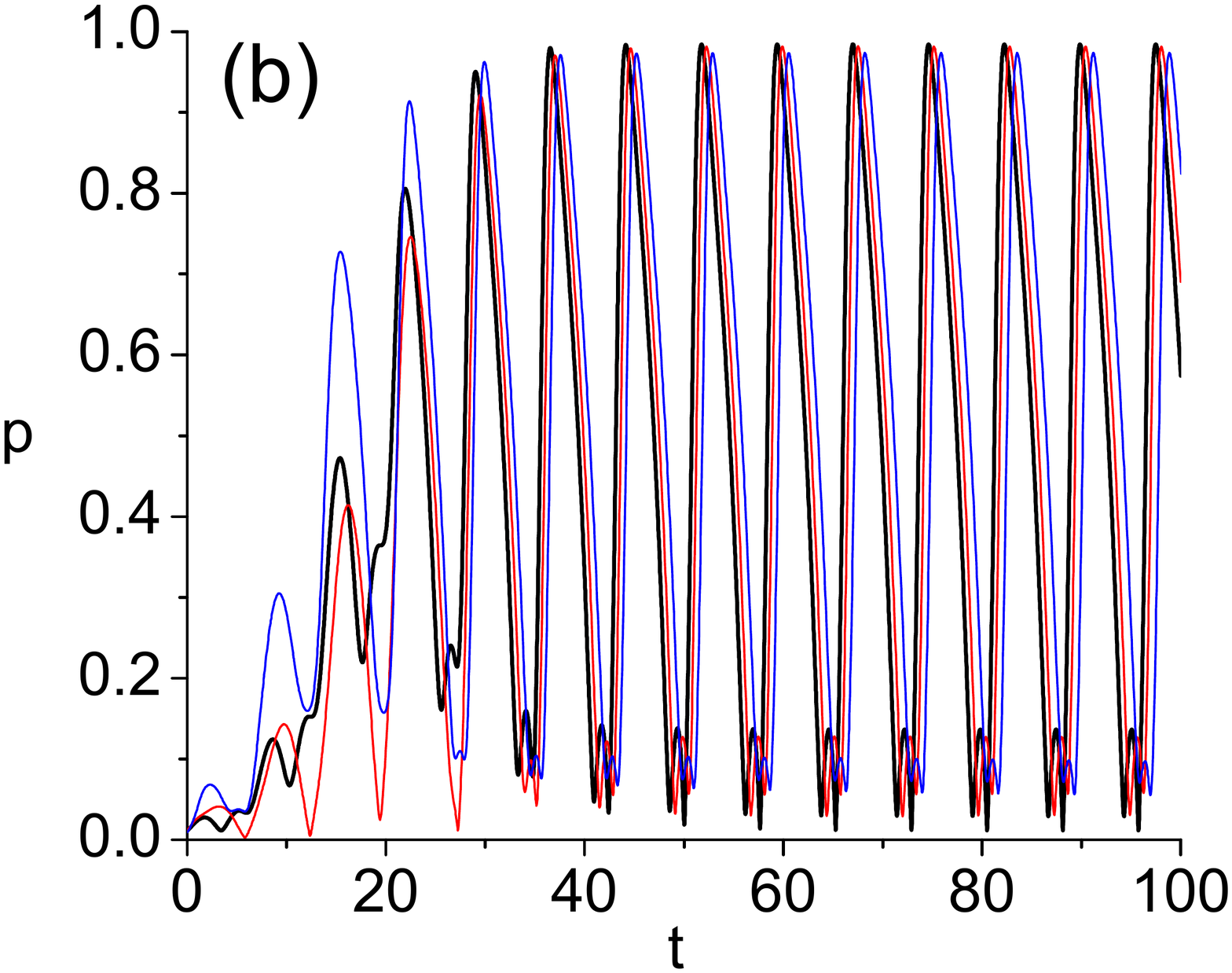} 
	\includegraphics[width=0.32\linewidth]{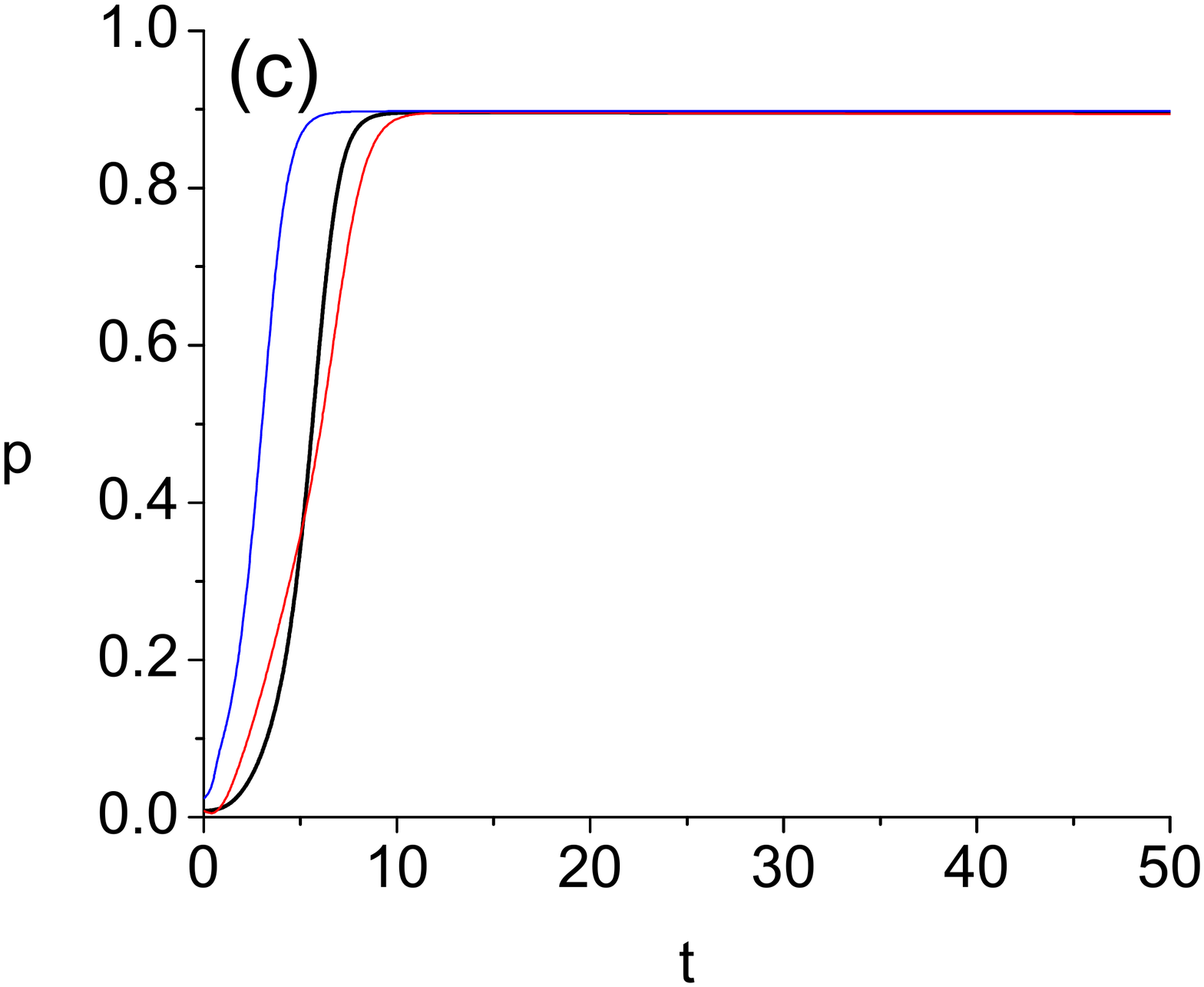} 
	\caption{(a) Order parameter as a function of the coupling constant for the uniform angular distribution of natural frequencies and $|\vec\omega_i|=1$. The
		continuous curve shows the numerical integration from Eq. (\ref{deltaf}) and the dots the value obtained by integrating the exact equations of motion (\ref{3dkura}). 
		For $K < 1.5$ the system converges to a limit cycle. The dashed line is the quadratic approximation, Eq.(\ref{quad1}). 
		(b) $p(t)$ for $K=1$; (c) $p(t)$ for $K=2$. Thin red and blue line shows the ansatz solution using our approach and that from \cite{chandra2019complexity}, respectively.} 
	\label{fig4}
\end{figure*}

For a delta function distribution $G(\vec{\omega}) = \frac{\delta(\omega-\omega_0)}{4\pi \omega^2}$ we find
\begin{equation}
	p = \int_0^{\pi/2} \rho_z(\bar{K},\theta) \sin \theta d \theta \equiv F(\bar{K})
	\label{deltaf}
\end{equation} 
where $\rho_z = - \alpha \sin \theta + \gamma \cos \theta$ depends only on $\bar{K}=  K p /\omega_0$ and $\theta$.  Once
the function $F(\bar{K})$ is calculated numerically, $p(K)$ can be computed in parametric form for $\omega_0=1$ as
$(p=F(x), K=x/F(x))$. The result is shown in Fig. \ref{fig4}. Equilibrium is found only for $K > 1.5$. For $K<1.5$ the
system converges to a limit cycle, which is well described both by the exact and reduced equations of motion. For $K=2$
our solution is slightly closer to the exact, whereas Chandra's is advanced, reaching equilibrium before the exact. In both 
cases the solutions present fluctuations that depend on the (random) initial conditions: every time the equations are integrated
a slightly different curve is obtained.

For small $K$ we can obtain analytical expressions expanding Eqs. (\ref{anasol})  to second order. We find 
$\alpha \approx  -\bar{K}^2 \sin \theta \cos \theta$, $\beta \approx -\bar{K} \sin \theta$ and $\gamma \approx 1 - (\bar{K}^2/2) \sin^2 \theta$. This gives 
\begin{equation}
	p = \frac{1}{2} + \frac{K^2 p^2}{8 \omega_0^2} \approx  \frac{1}{2} + \frac{K^2}{32 \omega_0^2}.
	\label{quad1}
\end{equation}
In the limit of very large coupling $\bar{K} \rightarrow \infty$, on the other hand, we get $\alpha \approx  - \sin \theta$, $\beta \approx - \sin \theta / \bar{K}$ and $\gamma \approx \cos \theta$,  which gives 
\begin{equation}
	p= \int_0^{\pi/2} (- \alpha \sin \theta + \gamma \cos \theta) \sin \theta d \theta 
	= \int_0^{\pi/2}  \sin \theta d \theta = 1.
\end{equation}

\begin{figure*}
	\centering 
	\includegraphics[width=0.45\linewidth]{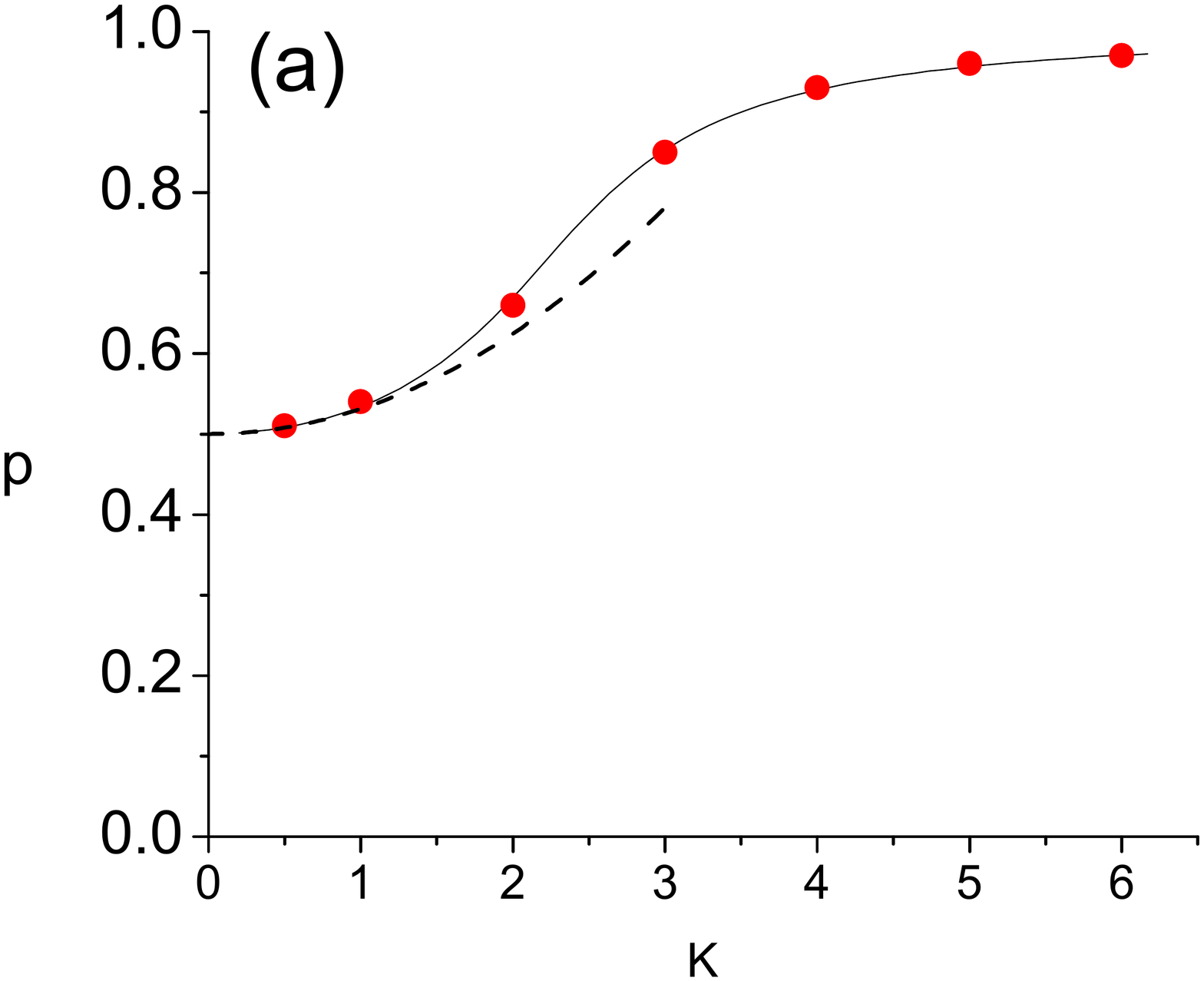} 
	\includegraphics[width=0.45\linewidth]{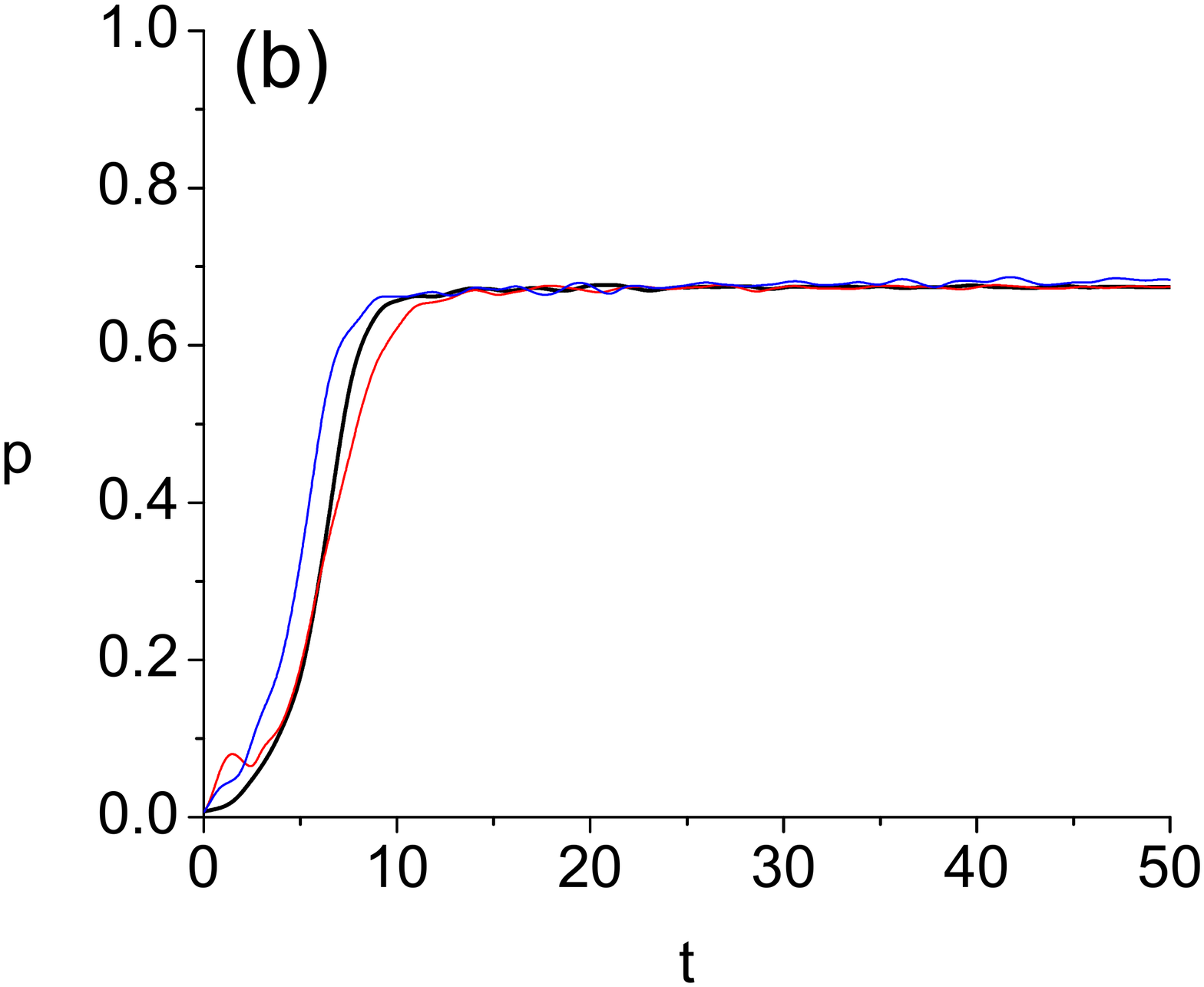} 
	\caption{(a) Order parameter as a function of the coupling constant for the Gaussian distribution of natural frequencies with $\Delta=1$. 
		The continuous curve shows the numerical integration from Eq. (\ref{gaussf}) and the dots the value obtained by integrating the
		exact equations of motion (\ref{3dkura}). The dashed line is the quadratic approximation, Eq.(\ref{quadratic}). 
		(b) $p(t)$ for $K=2$. Thin red and blue line shows the ansatz solution using our approach and that from \cite{chandra2019complexity}, respectively.} 
	\label{fig5}
\end{figure*}

For the Gaussian distribution $	G(\vec\omega)  = \frac{\sqrt{2}}{\sqrt{\pi}\Delta^3 4 \pi } e^{-\omega^2/ 2 \Delta^2}$ we have
\begin{equation}
	p = \frac{\sqrt{2}}{\sqrt{\pi}\Delta^3} \int_0^{\pi/2} d \theta \, \sin \theta  \int_0^{\infty} d \omega \, \omega^2 \, e^{-\omega^2/ 2 \Delta^2} \rho_z(Kp/\omega,\theta) .
\end{equation} 
Changing the integration variable to $\Omega = \omega/Kp$ and setting $\Delta=1$ we obtain
\begin{equation}
	p = \frac{\sqrt{2}}{\sqrt{\pi}} K^3 p^3 \int_0^{\pi/2} d \theta \, \sin \theta  \int_0^{\infty} d \Omega \, \Omega^2 \, e^{-\Omega^2 p^2 K^2/ 2 } \rho_z(1/\Omega,\theta)  \equiv H(Kp).
	\label{gaussf}
\end{equation} 
Once again the function $H(x)$ can be computed numerically and the parametric curve is given by $(p=H(x), K = x/H(x))$. The result is shown in Fig. \ref{fig5}. For small $K$ we obtain
\begin{equation}
	p = \frac{1}{2} + \frac{K^2 p^2}{8 \Delta^2} \approx  \frac{1}{2} + \frac{K^2}{32 \Delta^2}.
	\label{quadratic}
\end{equation}

% %%%%%%%%%%%%%%%%%%%%%%%%%%%%%%%%%%%%%%%%%%%%%%%%%%%%
% %%%%%%%%%%%%%%%%%%%%%%%%%%%%%%%%%%%%%%%%%%%%%%%%%%%%
\section{Conclusions}

In this paper we studied the problem of dimensional reduction for the Kuramoto model in 2 and 3 dimensions. In 2D we
used the ansatz proposed by Ott and Antonsen \cite{Ott2008} and solved the equations for the ansatz parameters at
equilibrium for arbitrary distributions of natural frequencies, re-deriving Kuramoto expressions near the bifurcation
point. 

We extended the ansatz to 3D expanding the distribution function in spherical harmonics. The resulting function differs
from that obtained by Chandra et al \cite{chandra2019complexity} and satisfies the continuity equation only if the
coupling term is modified. Our ansatz is connected to a especific choice of the coupling between the oscillators. This modification, however, is only in its radial component, which does not affect the
exact equations of motion. The interpretation of the approximate density function is very similar to that in 2D: for $\rho 
\rightarrow 0$ it describes a uniform distribution over the sphere and for $\rho \rightarrow 1$ it gives a delta function
at position $(\Theta, \Phi)$. The continuity equation leads to a vector equation for the ansatz parameters that is exactly
that obtained in \cite{chandra2019complexity}. However, the connection between the ansatz and the order parameter 
differs and it is simpler in our approach. The relationship we find is the natural extension of the 2D case, where $\vec{p}$ is the integral of $\vec{\rho}$ 
over $G(\vec{\omega})$, whereas that found in \cite{chandra2019complexity} is more complicated. This might facilitate the
analysis of more general systems where the coupling is a full matrix ${\mathbf K}$ or in the presence of external forces.

As applications we consider two types of $G(\vec{\omega})$ consisting of uniform angular distributions with delta or
Gaussian distribution of frequency module. In both cases the ansatz equations describe very well the full system and
show the first order transition behavior as expected. Curiously, the case of identical oscillators, although agrees
exactly at equilibrium, shows a delay in the dynamics, as shown in figure \ref{fig2}. Among the examples we explored,
this was the only case where the ansatz in \cite{chandra2019complexity} was more accurate than ours.

The complete dimensional reduction is only achieved for the case of identical oscillators. For all
other cases we have to solve Eq.(\ref{eqm1f}) together with (\ref{eqp3d}). However, as pointed out in
\cite{chandra2019complexity}, the latter can be solved by Monte Carlo methods, sampling the distribution
$G(\vec{\omega})$, which converges fast for most distributions of interest.

Finally, we remark that our ansatz points to an alternative way to treat the Kuramoto model in higher dimensions. It allows us
to easily obtain the dynamics of the order parameter and study the behavior at equilibrium for different
distributions of oscillator's natural frequencies.

\begin{acknowledgments}
	This work was partly supported by FAPESP, grants 2019/2027-5 (MAMA), 2019/24068-0 (ANDB), 
	 2016/01343‐7 (ICTP‐SAIFR FAPESP) and CNPq, grant 301082/2019‐7 (MAMA). We would like to thank
	 Alberto Saa and Jose A. Brum for suggestions and careful reading of this manuscript.
\end{acknowledgments}

\appendix

\section{Derivation of the dynamics of $\rho$ in 2D}
\label{app1}

Rewriting Eq.(\ref{contid1}) explicitly as 
\begin{equation}
\frac{\partial f}{\partial \rho} \dot{\rho} + \frac{\partial f}{\partial \Phi} \dot{\Phi}  + (\omega+q_\phi) \frac{\partial f}{\partial \phi} - f q_r = 0
\end{equation}
and replacing the derivatives we find, after multiplying by $D^2$,
\begin{eqnarray}
-2[2\rho - (1+\rho^2) \cos \xi] \, \dot{\rho} - 2 \rho_\phi (1-\rho^2 ) \, \dot{\Phi} +  \nonumber \\
2\omega (1-\rho^2) \rho_\phi + 2(1-\rho^2)(\rho_\phi q_\phi + \rho_r q_r) - (1-\rho^4) q_r = 0
\label{eqm}
\end{eqnarray}

Using  $\vec{q} = {\mathbf K} \vec{p}$ and the relations $\rho_\phi = \vec{\rho} \cdot \hat{\phi} = -\rho \hat{r} \cdot
\hat{\Phi}$, $\rho_r = \vec{\rho} \cdot \hat{r}$ and $\omega \rho_\phi = - \hat{r} \cdot (\vec{\omega} \times
\vec{\rho})$ we can greatly simplify Eq.(\ref{eqm}) to 
\begin{eqnarray}
	2(1-\rho^2)\hat{r}\cdot \left[ \vec{\rho} \times \vec{\omega} + \rho \dot{\Phi} \hat{\Phi} + \dot{\rho} \hat{\rho} \frac{(1+\rho^2)}{1-\rho^2} - \frac{1}{2}(1+\rho^2) {\mathbf K} \vec{p}\right] - 4 \rho \dot{\rho} + 2(1-\rho^2) \vec{\rho} \cdot {\mathbf K} \vec{p}  = 0. \nonumber
\end{eqnarray}

Adding and subtracting $\dot{\rho} \hat{\rho}$ inside the square brackets we obtain
\begin{eqnarray}
	2(1-\rho^2)\hat{r}\cdot \left[ \vec{\rho} \times \vec{\omega} + \dot{\vec{\rho}} + \dot{\rho} \hat{\rho} \frac{(2\rho^2)}{1-\rho^2} - \frac{1}{2}(1+\rho^2){\mathbf K}  \vec{p}\right] + 4 \left[\frac{1}{2} (1-\rho^2)  ({\mathbf K} \vec{p} \cdot \vec{\rho}) - \rho \dot{\rho}  \right] = 0 \nonumber.
\end{eqnarray}

Since this equation must hold for all values of $\hat{r}$ each term in the brackets must be zero:
\begin{eqnarray}
	\dot{\vec{\rho}}  &=&  \vec{\omega} \times \vec{\rho} - \dot{\rho} \hat{\rho} \frac{(2\rho^2)}{1-\rho^2} + \frac{1}{2}(1+\rho^2) {\mathbf K} \vec{p} \\
	\dot{\rho} &=& \frac{1}{2} (1-\rho^2)  ({\mathbf K} \vec{p} \cdot \hat{\rho}) . \label{eqmm}
\end{eqnarray}

Taking the scalar product of the first of these equations with $\vec{\rho}$ and noting that $\vec{\rho} \cdot \dot{\vec{\rho}} = \rho \dot{\rho}$ 
we can check that the second equation is recovered. Therefore the equations are compatible and we can replace $\dot{\rho}$ given by
the second into the first equation to finally obtain
\begin{eqnarray}
	\dot{\vec{\rho}}  =  \vec{\omega} \times \vec{\rho} + \frac{1}{2}(1+\rho^2) {\mathbf K} \vec{p} -  ({\mathbf K}  \vec{p} \cdot \vec{\rho})  \vec{\rho}.
\end{eqnarray}
%

%%%%%%%%%%%%%%%%%%%%%%%%%%%%%%%%%%%%%%%%%%%%%%%%%%%%%%%%%%%%%%%%%%%%%%
\section{Derivation of Eq.(\ref{fv4})}
\label{app2}

From Eq.(\ref{timederiv}) we obtain
\begin{equation}
	\begin{split}
		D^{\frac{5}{2}}\frac{\del f}{\del t}=\left[\cos\xi(3+\rho^2)-\rho(5-\rho^2)\right]\dot\rho 
		+ 3\rho(1-\rho^2) \hat{r} \cdot [\dot{\Theta} \hat{\Theta}+ \sin \Theta \dot{\Phi}\hat{\Phi}]  
	\end{split}
\end{equation}

Using eq.(\ref{fv1}) and eqs. (\ref{der3})-(\ref{der1}) we write
\begin{equation}
	\begin{split}
		D^{\frac{5}{2}}  \nabla(f\vec v) &= -2(1-\rho^2)q_r D + (\omega_\phi + q_\theta)3\rho(1-\rho^2) \frac{ \del\cos\xi}{\del\theta}  
		+ \frac{1}{\sin\theta}(-\omega_\theta +q_\phi) 3\rho(1-\rho^2) \frac{\del\cos\xi}{\del\phi} \\
		&= 3\rho(1-\rho^2) \left[\omega_\phi \frac{ \del\cos\xi}{\del\theta} - \frac{\omega_\theta}{\sin\theta} 
		\frac{\del\cos\xi}{\del\phi} \right] \\
		&+ \left[-2(1-\rho^2)q_r D + 3\rho(1-\rho^2) q_\theta \frac{ \del\cos\xi}{\del\theta}  
		+ 3\rho(1-\rho^2) \frac{q_\phi}{\sin\theta} 
		\frac{\del\cos\xi}{\del\phi} \right].
	\end{split}
	\label{fv2}
\end{equation}

The first line contains terms that are independent of the coupling vector $\vec{q}$. They can be simplified as follows:
\begin{equation}
	\begin{split}
		\omega_\phi \frac{ \del\cos\xi}{\del\theta} - \frac{\omega_\theta}{\sin\theta}  \frac{\del\cos\xi}{\del\phi} =
		\frac{1}{\rho}(\omega_\phi \rho_\theta - \omega_\theta \rho_\phi) = 	\hat{r} \cdot (\hat{\rho} \times \vec{\omega}).
	\end{split}
\end{equation}
Finally, using eqs. (\ref{der3})-(\ref{der1}), the terms containing $\vec{q}$ can be written as
\begin{equation}
	\begin{split}
		-2(1-\rho^2)q_r D + 3(1-\rho^2)(q_\theta \rho_\phi + q_\phi \rho_\theta) \\
		= (1-\rho^2) \left[-2 q_r(1+\rho^2) + 4 q_r \rho_r + 3 q_\theta \rho_\theta + 3 q_\phi \rho_\phi\right].
	\end{split}
	\label{vf3}
\end{equation}

%%%%%%%%%%%%%%%%%%%%%%%%%%%%%%%%%%%%%%%%%%%%%%%%%%%%%%%%%%%%%%
\section{Derivation of the dynamics of $\rho$ in 3D}
\label{app3}

To simplify Eq.(\ref{eqmot}) we first add 
$\dot{\rho} \hat{\rho}$ to the middle term of the first line to obtain $3\rho(1-\rho^2)  \hat{r} \cdot \dot{\vec{\rho}}$. We also subtract an 
equal term to keep the equation unaltered. This gives
\begin{equation}
	\begin{split}
		3(1-\rho^2) \hat{r} \cdot [ \vec{\omega} \times \vec{\rho} - \dot{\vec{\rho}}]  + \dot{\rho} \left[\rho(5-\rho^2) - 4 \rho^2 \cos\xi \right] \\
		- (1-\rho^2)\left[-2 \beta (1+\rho^2) + 3 \vec{\rho} ({\mathbf K} \cdot \vec{p}) - \frac{3}{2} (1+\rho^2) \hat{r} \cdot  ({\mathbf K} \cdot \vec{p}) + 4 \beta \hat{r} \cdot \vec{\rho} \right]=0
	\end{split}
\end{equation}    
Next we put all terms containing the angular coordinates  $\hat{r}$ together. We obtain
\begin{equation}
	\begin{split}
		3(1-\rho^2) \hat{r} \cdot [ \vec{\omega} \times \vec{\rho} - \dot{\vec{\rho}} + \frac{1}{2}(1+\rho^2) ({\mathbf K} \cdot \vec{p}) 
		-\frac{4 \beta }{3} \vec{\rho} - \frac{4}{3(1-\rho^2)} \rho \dot{\rho} \vec{\rho}]  \\
		+ \dot{\rho} \rho (5-\rho^2)  - (1-\rho^2) \left[3 \vec{\rho} \cdot ({\mathbf K} \cdot \vec{p})   - 2 \beta (1+\rho^2)\right] =0
	\end{split}
\end{equation}    
This implies that the following equations must be satisfied simultaneously:
\begin{equation}
	\begin{split}
		\dot{\vec{\rho}} = \vec{\omega} \times \vec{\rho}  + \frac{1}{2}(1+\rho^2) ({\mathbf K} \cdot \vec{p}) 
		-\frac{4 \beta }{3}  \vec{\rho} - \frac{4}{3(1-\rho^2)} \rho \dot{\rho} \vec{\rho}
	\end{split}
	\label{eqm1}
\end{equation}    
and
\begin{equation}
	\begin{split}
		\dot{\rho} \rho (5-\rho^2)  = (1-\rho^2) \left[3 \vec{\rho} \cdot ({\mathbf K} \cdot \vec{p})   - 2\beta (1+\rho^2)\right] 
	\end{split}
	\label{eqm2}
\end{equation}    

Taking the scalar product of  (\ref{eqm1}) with $\vec{\rho}$ and noting that $\vec{\rho} \cdot \dot{\vec{\rho}} = \rho \dot{\rho}$ we find that equations (\ref{eqm1}) and (\ref{eqm2}) are compatible only if
\begin{equation}
	\beta = \frac{\vec{\rho} \cdot ({\mathbf K} \cdot \vec{p}) }{4}.
	\label{betaC}
\end{equation}
With this condition (\ref{eqm1}) and (\ref{eqm2}) become
\begin{equation}
	\begin{split}
		\dot{\vec{\rho}} = \vec{\omega} \times \vec{\rho}  + \frac{1}{2} (1+\rho^2) ({\mathbf K} \cdot \vec{p})  -  [\vec{\rho} \cdot ({\mathbf K} \cdot \vec{p}) ] \vec{\rho} 
	\end{split}
	\label{eqm1fC}
\end{equation}   
and
\begin{equation}
	\begin{split}
		\dot{\rho}  = \frac{1}{2}(1-\rho^2)  [\hat{\rho} \cdot ({\mathbf K} \cdot \vec{p}) ].
	\end{split}
\end{equation}    
%

%%%%%%%%%%%%%%%%%%%%%%%%%%%%%%%%%%%%%%%%%%%%%%%%%%%%%
\clearpage 
\newpage

\end{document}